# A High-Performance InAs/GaSb Core-Shell Nanowire Line-Tunneling TFET: An Atomistic Mode-Space NEGF Study

A. Afzalian, G. Doornbos, T.-M. Shen, M. Passlack, *Fellow, IEEE*, and J. Wu

*Abstract*—Using a tight-binding mode-space NEGF technique, we explore the essential physics, design and performance potential of the III-V core-shell (CS) nanowire (NW) heterojunction TFET. The CS TFET "line-tunneling" current increases significantly with the core diameter $d_C$ and outperforms the best III-V axial "point-tunneling" NW heterojunction TFET $I_{ON}$ by up to 6× for $d_C$ = 6.6 nm. Reaching such a high level of current at low supply voltage, however, requires and involves specific and sometime unanticipated optimizations and physics that are thoroughly investigated here. In spite of the commonly accepted view, we also show and explain the weak gate-length dependency observed for the line-tunneling current in a III-V TFET. We further investigate the effect of electron-phonon scattering and discrete dopant impurity band tails on optimized CS NW TFETs. Including those non-idealities, the CS-TFET inverter performance significantly outperforms that of the axial TFETs. The low-power (LP) $V_{DD}$ = 0.35V CS-inverter delay is comparable to that of the high-performance (HP) Si CMOS using $V_{DD}$ = 0.55V, which shows promise for a LP TFET technology with HP speed.

## I. INTRODUCTION

Based on band-to-band tunneling (BTBT) through gate modulation of reverse-biased PN junctions, the tunneling field-effect transistor (TFET) is a promising candidate as building block to reduce the power consumption in electronic circuits, owing to its ability to reach inverse subthreshold slope (SS) below the thermal limit (60mV/dec at room temperature) [1]. Due to its principle of operation that is BTBT based, the TFET also intrinsically features a lower drive current than a MOSFET. Its current versus gate-voltage ($I_D(V_G)$) characteristics show slow increase, saturation, and even decrease in the on state [2]. As of today, TFETs have shown drive-current levels compatible for low-power (LP) CMOS applications but cannot reach the higher drive-current level requested by the more demanding high-performance (HP) circuits [3]. Finding a combination of architecture and materials that allows for an on-current level compatible with HP applications but at reduced power supply voltage ($V_{DD}$) and power consumption is, however, of paramount interest for future CMOS technology nodes.

To increase the current drive, TFET architectures with the band-to-band tunneling (BTBT) current aligned with the gate-induced electric field, the so-called "line-tunneling" (LT) TFETs (Fig. 1a), have been proposed to replace "point-tunneling" (PT) TFETs, in which BTBT occurs along the direction perpendicular to the gate-induced electric field [4]. LT TFETs have been studied both theoretically [5,6] and experimentally [7,8] in group IV material systems. They have shown improved current drive vs. their PT TFETs counterpart, although the current drive was typically below 10µA/µm and too low for CMOS applications [9]. Recently the LT-TFET concept was applied to two-dimensional (2D) material systems [10,11]. Theoretical atomistic DFT-NEGF models predict a drive-current level of 75µA/µm [10], i.e. closer to LP application requirements and somewhat comparable with III-V PT TFETs.

To enhance BTBT and increase $I_{ON}$, III-V semiconductor based TFETs have indeed been found very attractive, since III-V materials can provide small tunneling masses and heterojunctions that present favorable staggered- or broken-gap alignments. Possibility of drive-current levels compatible for LP circuits, i.e. of the order of 1 or a few 100 µA/µm, have been predicted theoretically and demonstrated experimentally in various III-V materials, such as InAs/GaSb or InGaAs/GaAsSb III-V heterojunction PT-TFET technologies [12-15].

There is little report on the III-V LT TFETs. III-V LT TFETs fabricated by top-down [16,17] and bottom-up approaches [18] were studied experimentally. The device dimensions were µm size and suffered from large contact resistances. In [16], semi-classical simulations on µm-size devices were also performed and have predicted large on-current levels for the structure. As of today, there is no report of the physics, design and the fundamental performance limit of such a technology at scaled dimensions, which is the scope of this paper. As a case of study, we explore here a core-shell III-V heterojunction nanowire TFET, which is similar, but at scaled dimensions, to the device fabricated in [18]. Fabrication process by a bottom-up vertical approach and experimental results can also be found in [19]. We report the first atomistic quantum-transport modeling study of the III-V InAs/GaSb GAA NW CS TFET, shedding light on its physics, design and fundamental performance limit. We demonstrate, for the 1$^{st}$ time, a scaled TFET technology potentially suitable for HP applications.

In section II, our mode-space (MS) tight-binding (TB) NEGF modeling approach is presented, including its extension to dissipative transport and the algorithm optimizations to allow for the atomistic simulation of 10 nm diameter NW TFETs beyond the ballistic approximation. We then apply our simulation method to III-V CS NW heterojunction TFETs. In section III, the essential physics, design and performance potential of the ideal ballistic CS devices are studied. In section IV, we investigate the effect of electron-phonon scattering and discrete dopant-impurity band tails on optimized CS TFETs, assessing the impact of these fundamental sources of non-ideality on steep swing and performance.

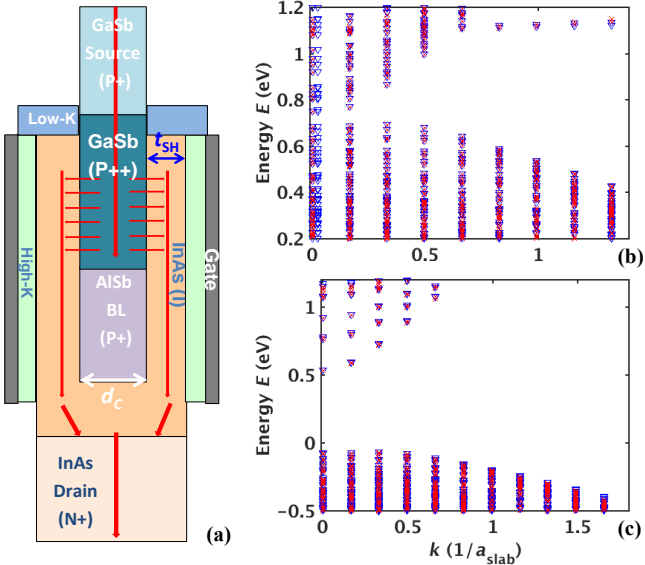

Fig. 1. (a) Schematic view of an optimized GaSb/InAs core-shell (CS) NW GAA PIN nTFET with core diameter $d_C$, and shell thickness $t_{SH}$. The total diameter $d = d_C + 2 \times t_{SH}$. The LT-current flow in the device in on-state is schematically represented by red arrows. E-k dispersion computed from the original (RS) TB models (▼) and from optimized (cleaned) MS bases (x) of (b) a GaSb core / InAs shell NW slab of a CS device with a $d$ = 10.2 nm, $d_C$ = 6.6 nm and $t_{SH}$ = 1.8nm ($C_{6.6}S_{1.8}$) and (c) a $d$ = 6.6 nm InAs NW slab.

## II. DEVICE STRUCTURE AND TB MS NEGF METHOD

Fig. 1a shows a schematic of a simulated CS InAs/GaSb GAA NW nTFET and introduces the necessary definitions. Due to the presence of BTBT within a strongly confined core and shell heterojunction, full-band quantum-transport simulations, such as atomistic tight-binding (TB) simulations within the NEGF framework [2],[20],[21], are required to accurately assess the performance of a scaled III-V line-tunneling TFET NW. Due to their computational cost, dictated by the 3D geometry (the full 3D geometry of the NW need to be explicitly simulated) and the atomistic nature, these simulations are too expensive to afford at technology relevant dimensions using a real-space (RS) technique [2,15]. We therefore used an efficient and accurate mode-space (MS) tight-binding (TB) NEGF method that has shown million-atom simulation capability for the simulation of axial NW PT HTFETs [2,22].

The general mode-space NEGF theory has been described in detail elsewhere [23-28]. Some specific algorithm optimizations that enabled us to simulate in a dissipative and atomistic way, by far, the largest NW reported in the literature merit, however, attention. We will only cover here the minimum set of equations to cover those and refer the interested reader to the reference above for more details. Using the recursive Green's Function (RGF) algorithm [29], the simulation time increases with a power-3 law with respect to the number of atoms in the cross-sectional slab of the nanowire, i.e. a power-6 law with respect to its diameter, and NWs with diameter smaller than 3 to 4 nm are typically simulated with atomistic RS NEGF methods when scattering is included [30,31]. The CS NWs simulated here feature diameters as large as 10.2 nm. Even using a combination of MS, state-of-the-art atomistic simulator and the latest generation Intel Xeon processors, one such IV simulation takes about 1 week on 400 cores using 15 GB of RAM/core for the ballistic case and 4 weeks on 800 cores using 19 GB of RAM/core for the case including electron – phonon scattering. This was achieved after a thorough code optimization and using an adaptive grid with 800 energy points.

The equations for retarded ($G^R$), lesser ($G^<$) and greater ($G^>$) Green's functions read [32]:

$$G^R = \left(EI_N - H - \Sigma^R\right)^{-1}, \quad (1)$$

$$G^< = G^R \Sigma^< G^{R\dagger}, \quad (2)$$

$$G^> = G^R - G^{R\dagger} + G^< \quad (3)$$

$E$ is the scalar energy. $I_N$ the identity matrix (of rank N), $H$ the device Hamiltonian, and $\Sigma^{R,<}$ the retarded, lesser self-energies that include the interaction terms (e.g. with the semi-infinite leads $\Sigma_C^{R,<}$ and the electron – phonon scattering terms $\Sigma_S^{R,<}$) are matrices of rank $N$, the total number of atoms in the device × the number of orbitals/atoms.

To change from the original real space of size $N$ to the reduced mode space of size $n$ ($n < N$), a block-diagonal unitary transformation matrix $U$ of size $N \times n$ has to be constructed. Each sub-block $U_{xi}$ of $U$ is the transformation matrix for the $x_i^{th}$ slab of the device composed of the $n_{xi}$ chosen orthonormal basis eigenvectors $\{\Psi_{xi}\}$ in the $N_{xi}$ - dimensional slab. In matrix notation, any approximate MS quantity, e.g., the device Hamiltonian is expressed as:

$$h = U^\dagger H U \quad (4)$$

In the text we use capital letters for RS quantities and small letters for MS ones. By transforming each terms of the right-hand side of equation (1) and (2) using (3), we obtain the mode-space expression for $g^R$, $g^<$ and $g^>$:

$$g^R = \left(EI_n - h - \sigma^R\right)^{-1}, \quad (5)$$

$$g^< = g^R \sigma^< g^{R\dagger}, \quad (6)$$

$$g^> = g^R - g^{R\dagger} + g^< \quad (7)$$

For atomistic Hamiltonians, constructing the slab unitary transformation matrices, $U_{Xi}$, requires an optimization procedure to clean the unphysical modes that arise from selecting an incomplete subspace basis [2,20,22]. Fig. 1b-c show and compare to the RS model, an essentially unphysical-mode free band structure (BS) obtained from a cleaned MS basis for a GaSb core / InAs shell slab of a CS device with a total $d = 10.2$ nm, core diameter $d_C = 6.6$ nm and shell thickness $t_{SH} = 1.8$ nm, a $C_{6.6}S_{1.8}$ device, as well as for an InAs axial NW slab with $d = 6.6$ nm.

### A. Real-space vs. mode-space density integration:

The NEGF computed real-space electron ($n$) and hole ($p$) densities are requested by the Poisson equation solver to perform the self-consistent loop. The RS densities at coordinate $r_i$ (typically an atomic site and orbital number in an atomistic basis) are obtained by integrating the RS diagonal elements of the lesser and greater Green's functions:

$$n(r_i) = \frac{1}{2\pi} \int_{E_N(r_i)}^{\infty} -iG^<(r_i, r_i, E) dE \qquad (8)$$

$$p(r_i) = \frac{1}{2\pi} \int_{-\infty}^{E_N(r_i)} iG^>(r_i, r_i, E) dE \qquad (9)$$

$$E_N(r_i) = \frac{E_V(r_i) + E_C(r_i)}{2} + V(r_i) \qquad (10)$$

$E_V(r_i)$ is the valence band maximum, $E_C(r_i)$ the conduction band minimum and $E_N(r_i)$ the neutrality point (typically sets as the middle of the bandgap energy) including the local electrostatic potential energy $V(r_i)$ at coordinate $r_i$. Our calculation being performed in MS, $g^<$ and $g^>$ need to be transformed back to RS before computing $n$ and $p$. Taking as an example the electron concentration, a direct transcription of eq. (8) is achieved by first up-converting $g^<(E)$ to real space for each energy, $E$, using the reverse transformation operation that the one used in eq. (4):

$$G^< = U g^< U^\dagger \qquad (11)$$

Then, the integration in real space using eq. (8) is performed. The real-space integration method is general and ensure that the density integration uses the local $E_N(r_i)$ including the local potential for each atom. It does, however, require up-converting the lesser or greater Green's functions individually for each energy $E$. An alternative and faster method would consist of swapping the order of up-conversion and integration. In that case the densities are integrated in MS directly. Then the densities are obtained within a single up-conversion:

$$n = \frac{-i}{2\pi} U \{ \int_{\tilde{E}_N}^{\infty} g^<(E) dE \} U^\dagger \qquad (12)$$

This is, however, only possible if a MS neutrality point can be found. As the only remaining spatial information is the position of the slab $x_i$ (the remaining MS degrees of freedom are eigenmodes indices and a clear relation between those and the energy-subband information is typically lost by construction of the optimized unitary transformation matrix in a full-band atomistic representation), we need to find a common neutrality-Energy $\tilde{E}_N(x_i)$ definition that holds for every atom of the slab. This is possible, providing that the slab bandgap is sufficiently larger than the local variation of the neutrality point of individual atoms, using the averaged neutrality point of all the $N_{Xi}$ atoms×orbitals of the slab:

$$\tilde{E}_N(x_i) = \sum_{N_{Si}} \frac{E_N(r_i, r_i \in x_i)}{N_{Si}} \qquad (13)$$

In that case, despite the local potential variation on different atoms, the average neutrality point is sufficiently within the gap so that its exact location is inconsequential to the electron and hole total integrated values.

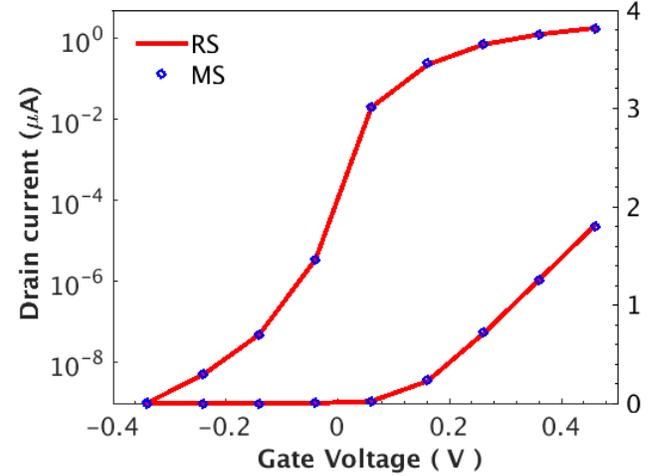

Fig. 2. $I_D(V_G)$ characteristics of a $C_3S_{1.2}$ NW CS nTFET computed from the RS TB model and from optimized (cleaned) MS bases. The relative MS to RS error is ≤ 1%. Typical MS to RS speedup are > 100×.

In the axial NW case, each slab is either composed of InAs or GaSb, and a bandgap of at least several 100 meV (Fig. 1c) is present, allowing for the faster MS integration of the density. It was verified that such an approach gives results in perfect agreement with RS NEGF [2,15]. In the CS device, at least part of the device features slabs that encompass both materials with a broken or close to broken slab bandgap (Fig. 1b). In that case, a neutrality point only exists locally and only the slower real-space density integration yields results in good agreement with the RS NEGF results (Fig. 2). Note that, for real-space densities, only the diagonal element of $G^{<,>}$ are needed. This is exploited in our algorithm to considerably fasten the up-conversion operation (11). Still, the RS-density integration scheme typically increases the simulation time by 1.5 to 1.8×, when compared to the faster MS-density integration method.

### B. Electron–phonon scattering:

Within the self-consistent Born approximation, the self-energy for the electron-phonon interaction is:

$$\Sigma_S^< = D^< G^< , \qquad (14)$$

with the free phonon Green's function $D^<$ [32]. Assuming the phonons stay in equilibrium, (14) can be written as:

$$\Sigma_S^<(r_i,r_j,E) = \int \frac{dq}{(2\pi)^3} e^{iq\cdot(r_i-r_j)} |M_q|^2$$
$$\times (N_q + \frac{1}{2} \pm \frac{1}{2}) G^<(r_i,r_j,E \pm \hbar\omega_q) \quad (15)$$

where $q$ and $\omega_q$ are the phonon wave vector and the corresponding angular frequency, $\hbar$ is the reduced Plank's constant, $N_q$ is the phonon occupation number. $M_q$ is the electron-phonon coupling matrix, which depends on the exact scattering mechanism.

In MS, to compute (15), it is possible to up-convert $g^<$ to RS using (11), compute $\Sigma_S^<$ using (15), then, finally compute $\sigma_S^<$ using (4) for further use. This seriously impact the speed and memory usage of the simulation due to the number of up and down conversions from MS to RS and the need to store in memory the very large RS matrices. To tackle the numerical burden of considering e-ph in such large NW devices, we therefore used a direct MS equivalent expression based on a form factor method [28]:

$$\sigma_{S,mn}^<(x_i,x_j,E) = \int \frac{dq}{(2\pi)^3} e^{iq_x(x_i-x_j)} |M_q|^2$$
$$\times \sum_{kl} (N_q + \frac{1}{2} \pm \frac{1}{2}) g_{kl}^<(x_i,x_j,E \pm \omega q)$$
$$\times F_{mn}^{kl}(x_i,x_j,q_t) \quad (16)$$

where $m$, $n$, $k$, $l$ are slab eigenmode indices, while $q_x$ and $q_t$ are the longitudinal and transversal components of the phonon wave vector $q$, respectively. To solve this equation, we need to do some approximations. In this work acoustic, optical and polar-optical phonons will be considered within a local approximation. That is, we only keep terms with $r_i = r_j$ in (15). Although the NEGF formalism provides a theoretical framework to consider non-local scattering, the local approximation has to be done to use the faster RGF algorithm and keep the simulation time and memory manageable in a device with realistic dimensions (e.g., see [2,30]). Acoustic and optical phonons tend to be local mechanisms as their $M_q$ term is not much $q$ dependent [2,28]. Polar-optical phonons, on the other hand, feature non-local components, but we only keep here the local (largest) interaction term [2]. Using the local approximation, the scattering self-energy $\Sigma_S^<$ is diagonal and the form factor $F$ simplifies to:

$$F_{mn}^{kl}(x_i) = \int \Psi_m^*(y,z;x_i) \Psi_n(y,z;x_i)$$
$$\times \Psi_k(y,z;x_i) \Psi_l^*(y,z;x_i) dydz \quad (17)$$

Still, this expression is a power-4 expression of the $x_i^{th}$ slab $n_{xi}$ MS basis vectors. Despite the reduction of the slab size from $N_{xi}$ = several 10 000 in RS to $n_{xi}$ = several 100 in MS, this expression rapidly becomes inapplicable in an atomistic framework. For the $d$ = 10.2 nm case, typical $n_{xi}$ values are ranging from 550 to 600, leading to the computation and application of more than 100 billion form factors per slab. To keep the problem tractable, we used an uncoupled-mode approximation for the computation of the self-energy, that is, we only keep the terms with $m=n$ and $k=l$ in (16) and (17). If the mode coupling between the modes is neglected, $g^<$ and $\sigma_S^<$ become diagonal [26]. In practice, even with mode coupling, $g_{kl}^<$ terms with $k \neq l$ are typically small compared to $g_{kk}^<$ ones. In addition, inspecting the form-factor equation (17), it can be seen from the orthonormality of the wave function that terms with $m \neq n$ and $k \neq l$ are small, due to the reduced wave-function overlap. As a result of these two facts, it is reasonable to expect that terms with $m \neq n$ and $k \neq l$ add little contribution to $\sigma_S^<$ and can be neglected. In [28], we have verified using effective-mass Hamiltonians that the uncoupled-mode approximation for the self-energy gives accurate results, even in case of strong mode coupling. A similar validation was performed for tigh-binding bases [33]. Note that the uncoupled-mode approximation is only used for the self-energy calculation, while the mode-coupling terms ( $g_{kl}^<$ terms $k \neq l$ that are not further multiplied with a reduced form factor term) are kept in (11) for the carrier density calculation [28,30].

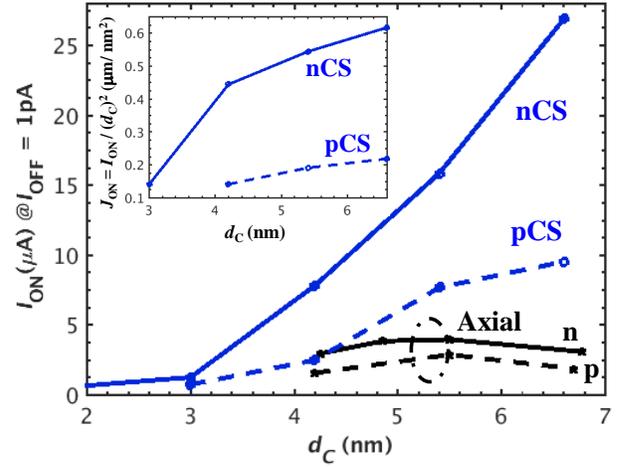

Fig. 3. Impact of (core) diameter on $I_{ON}$ (current per wire) of optimized ideal axial, and CS InAs/GaSb GAA NW n- and pTFETs. $I_{OFF}$ = 1 pA/wire, $V_D$ = 0.3 V. Gate oxide: 1.8 nm $Al_2O_3$ oxide. For the CS TFETs, the on-current density $J_{ON}$, i.e., the ratio of $I_{ON} / d_C^2$ is also shown (inset). $V_G$ = 0.3 V.

Because of the interdependence between $g^<$ and $\sigma^<$ (see (6) and (16)), it is common to add to the outer Poisson-NEGF loop, an inner loop that iterates (6) and (16). After a new potential has been computed and used to update the Hamiltonian of the system, one starts with an initial guess for $\sigma^<$. It can be $\sigma_C^<$ (the contact self-energy, i.e., $\sigma_S^<$ = 0, the ballistic case, used for the first iteration) or $\sigma_C^<$ + a $\sigma_S^<$ value computed from a previous Poisson-NEGF iteration and interpolated to the updated energy mesh of the new outer iteration [28]. (6) and (16) are then iterated up to convergence of the charge density. This approach typically requires many expensive NEGF iterations. We have verified that removing the inner loop (i.e., performing (6) and then (16) only once, using for (6) the interpolated $\sigma_S^<$ values from the previous Poisson iteration) is stable and accurately converges for the NW case. It leads to the fastest convergence (i.e. the less NEGF iterations to compute a bias point) in most cases. We used this approach here to fasten the computation while maintaining excellent accuracy.

## III. IDEAL CS HTFETs

### A. Physics and design:

Fig. 3 shows the dependence of the simulated ballistic on-current ($I_{ON}$) on $d_C$ for optimized InAs/GaSb heterojunction CS and axial n- and pTFETs. The CS TFET $I_{ON}$ per wire significantly increases with $d_C$ and outperforms the best axial TFET $I_{ON}$ by up to 6×. Reaching such a high level of current at $V_{DD} = 0.3V$, however, requires specific optimizations as we will describe below.

Fig. 4a and 5a show the shell conduction-band ($CB_{SH}$) and core valence-band ($VB_C$) edges along the channel direction, $x$, as well as the current spectrum $J(x, E)$ in off- and on-state, respectively, for a well-designed ideal $C_{5.4}S_{1.8}$ CS nTFET. In the off-state (Fig. 4a) $CB_{SH}$ is at a higher energy than $VB_C$ and the BTBT path from the core to the shell is closed. The core-channel-to-drain BTBT leakage current is suppressed by the use of a wider-bandgap lattice-matched AlSb blocking layer (BL). In the on-state (Fig. 5a), $CB_{SH}$ is pushed below $VB_C$ in the channel by the action of the gate electric field. A virtually 0 tunneling distance and a full transmission of the device transmission states (TS) can be achieved in a given energy range, enabling a high "line-tunneling" drive current.

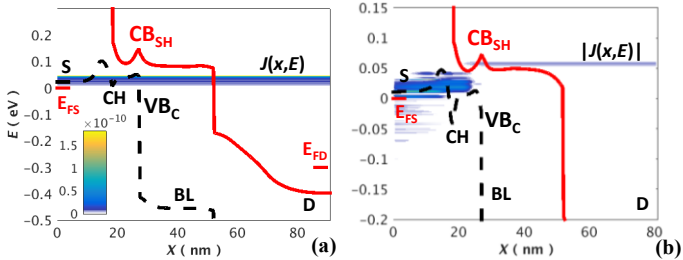

Fig. 4. Current spectrum $J(E,x)$ in A/eV (surface plot), as well as shell conduction band ($CB_{SH}$) (-) and core valence band ($VB_C$) (--) edges, along the channel direction $x$ for the optimized $d_C = 5.4$ nm $C_{5.4}S_{1.8}$ CS nTFET in off-state (a) under ballistic approximation and (b) with electron-phonon scattering (e-ph).

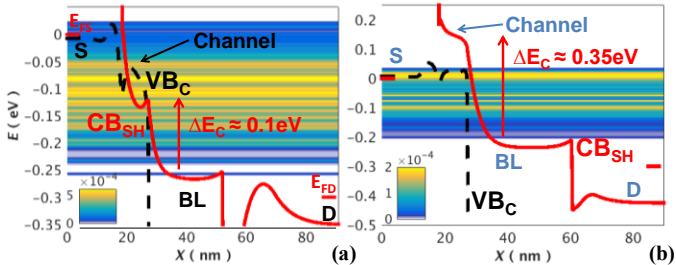

Fig. 5. Current spectrum $J(E,x)$ in A/eV, $CB_{SH}$ (-) and $VB_C$ (--) edges along $x$ for the a $d_C = 5.4$ nm $C_{5.4}S_{1.8}$ CS nTFET in on-state under ballistic approximation (a) with aligned channel and BL shell conduction bands using a dual work-function gate, (b) without $CB_{SH}$ band alignment.

The low level of on-current observed for $d_C < 4$ nm in a CS nTFET is caused by the combination of staggered InAs/GaSb effective bandgap (related to the strong quantum confinement (QC)) and core-channel depletion (CCD) by the gate field in narrow CS devices. CCD prevents carriers to populate the core-channel in on-state and strongly suppresses the inversion of $CB_{SH}$ and $VB_C$, hence line-tunneling. It needs to be mitigated by a high core-channel doping, $N_{CC}$, (Fig. 6). The smaller $d_C$, the larger the doping needs to be. For $d_C = 1.8$ nm, even with a doping concentration $> 1 \times 10^{21}$ cm$^{-3}$, i.e., typically above the solubility limit of these III-V materials, it was not possible to prevent CCD and achieve LT, so that the drive current remained low. At $d_C = 5.4$ and 6.6 nm, a $N_{CC} = 8 \times 10^{19}$ (see inset of Fig. 6) and $5 \times 10^{19}$ cm$^{-3}$, respectively, is optimal.

Similar causes hamper the performance of the CS pTFET at small $d_C$. Compared to the n-case, a larger $d_C$ is required to enable a strong line-tunneling current due to the larger effective bandgap (related to a stronger quantum confinement in the InAs core) observed at same diameter in the p-case.

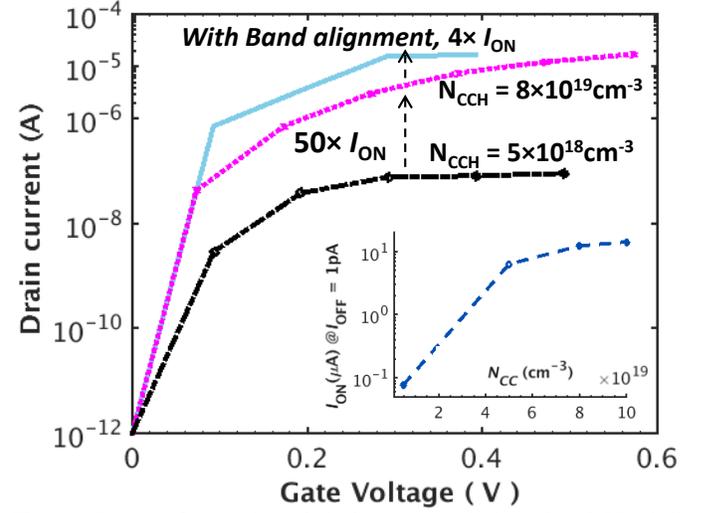

Fig. 6. Impact of core channel doping, $N_{CC}$, and channel and BL shell conduction bands alignment on $I_D(V_G)$ characteristics of an optimized InAs/GaSb CS GAA NW nTFET. The impact of $N_{CC}$ on $I_{ON}$ (current per wire) is also shown (Inset). Channel length $L= 9$ nm. $t_{SH} = 1.8$ nm. $d_C = 5.4$ nm. $V_D = 0.3$ V.

Further increasing $d_C$ above 4 nm increases the junction area and the number of available TS from source to drain without impacting the electrostatic control over the junction, hence results in a net increase of the drive current (Fig. 3). This is not the case for the axial TFETs, where an optimal diameter of 5.5 nm was found (see Fig. 3 and [15]). The inset of Fig. 3 shows $J_{ON}$, the ratio of $I_{ON} / d_C^2$, vs. $d_C$. For $d_C > 4$ nm, both for the CS n- and pTFETs, the drive current approximately scales with $d_C^2$, i.e., consistently with the increase of the number of available TS (a strict $d_C^2$ scaling would imply a constant $J_{ON}$, the fact that $J_{ON}$ further improves with $d_C$, especially in the n-case, is related to the reduction of the QC and core depletion with $d_C$).

The close to 3× smaller current density for the p-case, when compared to the n-case, is partially related to the inherent dissymmetry of the conduction and valence band density of states (DoS) of the III-V materials, as already well known for the axial TFET case (e.g., see [2,15,34]). In the CS TFET case, the p-current drive is further affected by the strong quantum confinement in the InAs core (as already mentionned above).

We expect that larger p-type current densities could be observed by further reducing this confinement, e.g., for even larger core diameters, by using a nanosheet rather than a NW for instance, or by finding a better-suited core material than InAs.

The source is doped to $5\times10^{19}$ cm$^{-3}$ in the nTFET case, and to $5\times10^{18}$ cm$^{-3}$ in the pTFET case to avoid strong degeneracy that is detrimental for SS of a TFET, similarly to what was found in the axial TFET case [2,34]. Also, as for the axial TFET, the CS drain doping results from a trade-off between on and off-current [2,15], and is close to $5\times10^{18}$ cm$^{-3}$ for the n- and close to $1\times10^{19}$ cm$^{-3}$ for the pTFET. To minimize on-current saturation effects and achieve well-saturated output characteristics, a doping level sufficient to degenerate the drain band edge is needed, while to limit the ambipolar current the drain doping cannot be further increased, and a lowly-doped region with a length in the range of 5 to 10 nm is used between the BL and drain region (Fig. 1a).

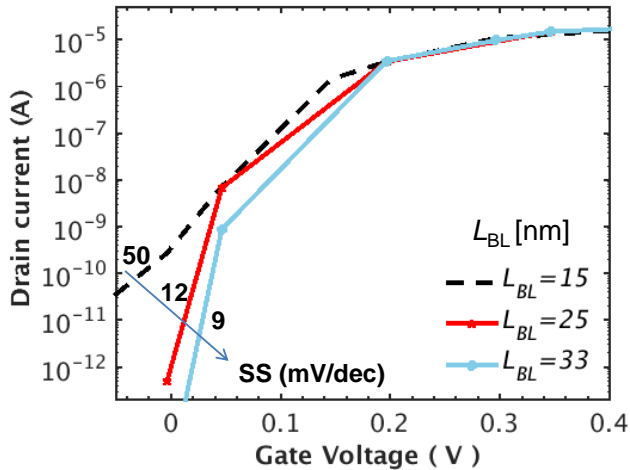

Fig. 7. Impact of barrier layer length $L_{BL}$ on $I_D(V_G)$ characteristics of an optimized InAs/GaSb CS GAA NW nTFET. $L$ = 9 nm. $t_{SH}$ = 1.8 nm. $d_C$ = 5.4 nm. $V_D$ = 0.3 V.

As mentioned above, the core channel is highly doped to prevent core depletion. The high-doping region, as well as a low-K dielectric spacer with a relative permittivity close to 4, extends for several nm at the source side to limit source depletion by the gate fringing field that is detrimental to the drive current in the on-state. One unwanted consequence of the high core doping in the channel is that the shell bands over the channel are shifted in energy compared to those of the shell that are physically above the undoped BL. The shift is in such a way that, without re-aligning the bands, point-edge conduction from the edge of the channel to the BL shell would occurs first and a large gate overdrive (typically > 0.6V) would be needed to achieve the desired high-drive LT current.

This is illustrated in Fig. 5 and 6 for a CS nTFET with $d_C$ = 5.4 nm and $N_{CC} = 8\times10^{19}$ cm$^{-3}$. Intrinsically (without doping and electrostatic effects), due to the different core materials, the InAs shell conduction band over the channel already presents a small energy shift of a few ten of meV with respect to that of above the BL. However, when the core-channel is doped while the barrier-layer core remains intrinsic, the InAs CB over the channel is further increased by several 100 of meV, so that the path for line tunneling is not opened at $V_G$ = 0.3V and only PT-tunneling current flows (Fig. 5b).

There are various ways to re-align the shell bands, all resulting in a similar and strong improvement of the drive current at $V_G$ = 0.3V, when compared to the non-aligned case (Fig. 6). One is simply to dope the barrier layer in a similar fashion than the core channel. For instance, in the $d_C$ = 5.4 nm nCS TFET of Fig. 6, by P-doping the core BL to $5\times10^{19}$ cm$^{-3}$, we can, with minimal impact on the off-state leakage, re-align the bands with less than 100 meV difference. This is sufficient to ensure that LT turns on around $V_G$ = 0.2V and $I_{ON}$ at $V_G$ = 0.3V is more than $4\times$ larger than that in the un-aligned case. Another possibility is to use a dual work-function gate. By having the work function of the gate that is physically over the channel 0.6eV lower than that over the BL, a band alignment and performance similar to those of the doped barrier-layer case are achieved (Fig 5a). A 3$^{rd}$ possibility is that of using a different shell material with an appropriate band-offset (e.g. using an InGaAs shell that we have assumed pseudomorphically strained) over the BL, or a combination of 2 or 3 of these methods. The 3rd option is also beneficial to reduce the leakage floor of the device by selecting a larger bandgap, higher effective-mass material in the shell of the BL as discussed next.

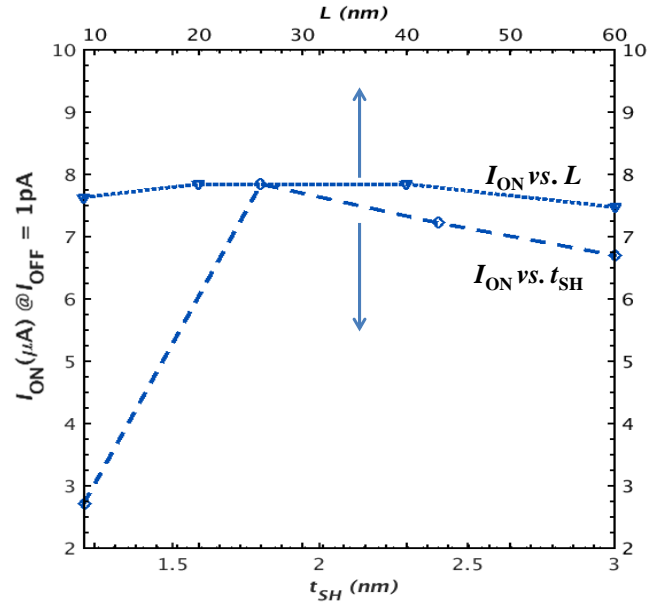

Fig. 8. Impact of shell thickness ($L$= 20 nm) and channel length ($t_{SH}$ = 1.8 nm) on $I_{ON}$ (current per wire) of optimized $d_C$ = 4.2 nm CS InAs/GaSb GAA NW nTFETs. $I_{OFF}$ = 1 pA/wire, $V_D$ = 0.3 V.

The impact of the BL length, $L_{BL}$, on the leakage current and SS is shown on Fig. 7 for the n-case. At $L_{BL}$ = 15 nm, the leakage features both an axial tunneling component through the BL VB$_C$ and a tunneling component under the shell BL CB. For longer $L_{BL}$ however the axial component becomes insignificant compared to the leakage current under the shell BL CB. As the InAs shell has a lower effective mass and bandgap than the BL core, a longer value $L_{BL} \geq 25$ nm is to be used to minimize shell related leakage below 1pA/NW. The situation is similar for the p-case, but a shorter $L_{BL} \geq 20$ nm can be used to minimize shell leakage below 1pA/NW, due to the larger effective masses in the GaSb shell. To further reduce $L_{BL}$ down to about 15 nm, larger effective masses and bandgap materials in the BL shell (e.g., InGaAs for the n-case, AlGaSb for the p-case) can be used. For even shorter $L_{BL}$ both BL shell and core material need to be changed. The fact that the BTBT LT drive current is determined by the channel core-shell material, while the off-state leakage is controlled by the BL core-shell material is an advantage of this structure. It opens, at least theoretically, many new possibilities and material combinations to optimize on and off-current independently. This is not the case in the axial PT-TFET where using an InGaAs channel or InGaAs drain barrier was shown to improve off-state but degrade the on-state [15].

The shell thickness optimization results from a trade-off between electrostatic and QC (Fig. 8). Increasing $t_{SH}$ does not increase the amount of available TS, while decreasing the electrostatic control over the tunneling junction. This yields a small decrease of the maximum achievable $I_{ON}$ for $t_{SH} > 1.8$ nm in the considered range. Decreasing $t_{SH}$ down to 1.2 nm results in a severe $I_{ON}$ degradation. This is related to a strong increase of $CB_{SH}$ and effective bandgap due to QC.

Next, we investigate the impact of the channel length $L$ (Fig. 8). The current drive was found to depend only weakly on $L$ (the current is rather source DoS limited in the considered range), and $L$ can be scaled down below 10 nm. Based on a simple geometrical argument -increasing $L$ increases the tunneling junction area- it is expected that the current scales linearly with the gate length in a LT-TFET. Such behavior was modelled analytically [5] and observed experimentally [7] in a Si/SiGe LT-TFET, where the current linearly scaled with $L$ up to several hundred of nm before it started to saturate. We note, however, that a current that increases with $L$ is against the basic laws of physics (e.g., Ohm law for the diffusive regime or Landauer formula in the ballistic case) and no one expect a good metallic conductor or even a MOS transistor to drive more current when increasing $L$. It can only be observed in cases where the transmission of the tunneling junction is so poor that it strongly limits the current of the device.

The problem is as follow. In the standard junction case, e.g., an axial tunneling junction where the junction area is the cross-section area, increasing the junction area directly increases the number of transmission states from source to drain to drive the current. In the LT-case, however, increasing $L$ has no impact on these TS, so that it can only increase the current by increasing the transmission probability (e.g., by opening a larger portion of the energy windows to LT) through the junction of the fixed number of device TS. Once increasing $L$ does not improve the transmission probability of individual TS, e.g., these are all close to full conduction already, the current can only saturate. Due to the very good transmission probability of the broken-gap low effective-mass InAs/GaSb system, this saturation effect happens for $L$ of a few nm only. We note that a similar prediction of the current not increasing for $L ≥ 20$ nm was made in a 2D broken-gap LT TFET [10].

To summarize, although increasing $L$ does increase the tunneling junction area, it does not increase the total number of TS that drive the current. The total number of TS and their relative degree of transmission (close to full transmission in the on-state of a well-designed CS device), not the junction area, defines the maximum achievable current.

*B. Performances:*

The 2017 release of the International Roadmap for Devices and Systems (IRDS) [9] specifies that from the year of production of 2030 and beyond, a vertical gate-all-around device architecture (VGAA) will be mainstream. This would allow for relaxing the device length and accommodate the necessary longer gate length (e.g., due to the barrier layer of the CS TFET) required for the III-V TFETs.

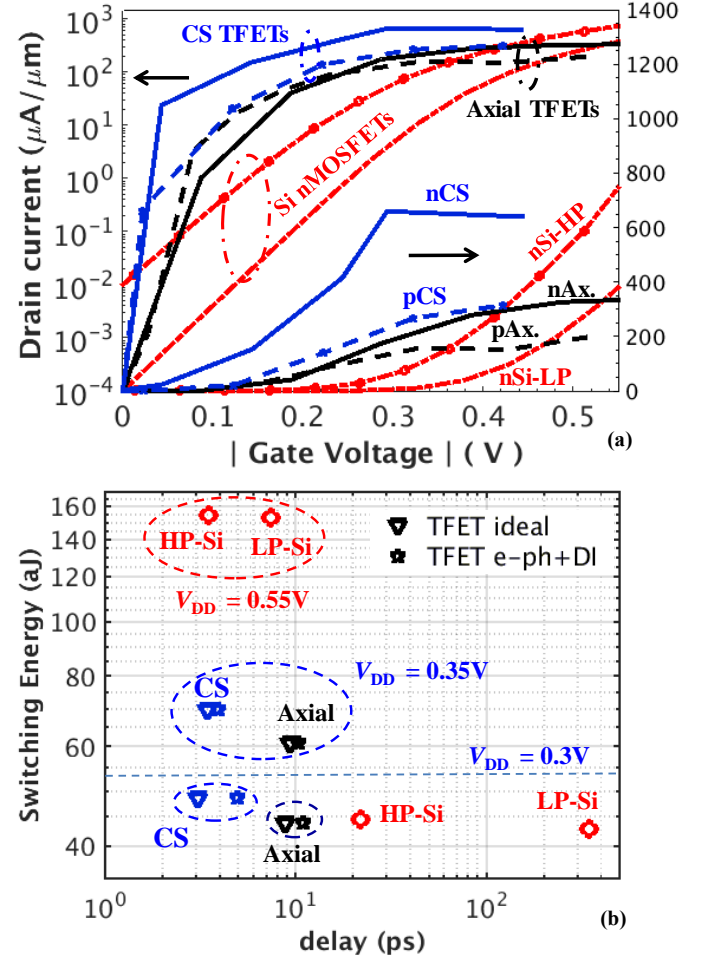

Fig. 9. (a) $I_D(V_G)$ characteristics of optimized Si nMOSFET at LP and HP leakage current specifications and $V_{DD} = 0.55$ V, and InAs/GaSb axial (Ax.) and CS GAA NW n- and pTFETs at LP current specification and $V_{DD} = 0.3$ V. (b) Averaged (between n and p) switching energy *vs.* delay (EDP) of a 5.5 track-high inverter cell for optimized Si MOSFETs, InAs/GaSb axial and CS GAA NW TFETs at various $V_{DD}$. To fit the 77 nm inverter cell height, 5 wires/device are used for the Si and axial NW TFET cases and 4 wires/device are used for the CS NW TFETs. The inverters are loaded with a 50 contact-gate pitches-long metal line [9] and the input capacitance of 3 identical inverter cells (Fan out of 3). The extrinsic capacitances of the cell layout are also included in the load capacitance. In the TFET cases, both ideal (ballistic) and non-ideal (e-ph scattering + DI bandtails) EDP's are shown. LP $I_{OFF}$ = 100 pA/μm. HP $I_{OFF}$ = 10 nA/μm.

Fig. 9a benchmarks the $I_D(V_G)$ characteristics of CS n- and pTFET VGAA designs vs. those of axial VGAA TFETs at $I_{OFF}$ = 100pA/μm and $V_{DD}$ = 0.3V and those of VGAA Si nMOSFETs at LP ($I_{OFF}$ = 100 pA/μm) and HP ($I_{OFF}$ = 10nA/μm) leakage current specifications with $V_{DD}$ = 0.55V. The current was normalized by the NW perimeters. For the CS TFETs the total diameter *d* including core and shell was used to compute the perimeter. For 2033, the IRDS targeted drive currents are 937 µA/µm for HP and 637 µA/µm for LP logic at $V_{DD}$ = 0.55 V. For comparison the Si HP and LP nMOS achieve 746 and 380 µA/µm, respectively, at $V_{DD}$ = 0.55 V. As the $I_{ON}$ values are calculated from a CV/I delay specification, the required current values for $V_{DD}$ = 0.3V can be estimated to be 511 µA/µm for HP and 347 µA/µm for LP. The LP n- and pCS TFETs achieve $I_{ON}$ = 659 and 239 µA/µm respectively at $V_{DD}$ = 0.3V, which for the n-case exceeds the IRDS $I_{ON}$ requirement for HP with both lower leakage (100× smaller) and lower active power consumption (due to the lower operating voltage). For the p-case, as discussed above, we expect that a value closer to the n-device could be achieved if a solution to reduce the confinement can be found. For comparison, the LP axial NW n- and pTFETs achieve $I_{ON}$ = 188 and 138 µA/µm, and the nSi NW MOSFET achieve 63 and 6 µA/µm for HP and LP $I_{OFF}$, respectively, at $V_{DD}$ = 0.3 V.

Fig. 9b benchmarks the CMOS-inverter energy and delay of the CS VGAA designs vs. those of axial VGAA TFETs at $I_{OFF}$ = 100pA/ µm and $V_{DD}$ = 0.3 and 0.35 V, as well as those of VGAA Si nMOSFETs at LP and HP $I_{OFF}$ with $V_{DD}$ = 0.3 and 0.55V. For 2033, IRDS predicts a metal half pitch of 7 nm and that 8-additional nm are required around the VGAA NWs to accommodate the gate stack and spacing between adjacent NWs [9]. Assuming a 5.5 Tracks standard cell, this yields a 77 nm-tall cell. Using these values and keeping the IRDS layout that uses 2 NWs in the width direction, we can fit at 1st order 10 Si NWs, 8 CS TFETs and 10 axial TFETs in one CMOS inverter cell.

As a result of its large current-drive, at $V_{DD}$ = 0.3V, the ideal CS-TFET CMOS inverter achieves the fastest delay, followed by that of the axial TFETs and the Si MOSFETs (Fig. 9b). For comparison, the $V_{DD}$ = 0.3V CS-TFET inverter delay is comparable to that of the HP Si CMOS using $V_{DD}$ = 0.55V, while the $V_{DD}$ = 0.3V axial-TFET inverter delay is comparable to that of the LP Si CMOS using $V_{DD}$ = 0.55V. In the TFET cases, owing to the lower supply voltage, the switching energy is reduced by more than 3×.

TABLE I
SUMMARY OF SI HP NMOSFET, AND INAS/GASB *AXIAL* AND *CS* GAA NW N- AND PTFET DESIGNS.

|  | CS III-V TFET | Axial III-V TFET | Si HP nMOSFET |
|---|---|---|---|
| $d_C$ (nm) | 6.6 | 5.5 | 6 |
| $t_{SH}$ (nm) | 1.8 | 0 | 0 |
| L (nm) | 9 | 20 | 12 |
| EOT (nm) | 0.7 | 0.7 | 0.7 |
| Orientation | [100] | [111] | [100] |
| $I_{OFF}$ (pA/µm) | 100 | 100 | 10000 |
| **Ideal Performances** | | | |
| $V_{DD}$ (V) | 0.3 | 0.3 | 0.3 / 0.55 |
| type | n / p | n / p | n |
| $I_{ON}$ (µA/µm) | 659 / 239 | 188 / 138 | 63/ 746 |
| *Performance degradation due to scattering (e-ph) + DI band tails* | | | |
| $\Delta I_{ON}$ @$V_{DD}$ = 0.3V | -38%/- | -14.8%/- | - |
| $\Delta I_{ON}$ @$V_{DD}$ = 0.35V | -13%/- | -4.4 %/- | - |

It is important to note that, contrary to the MOSFET case, it is not possible to trade power for delay and to further enhance the TFET speed by using a larger $V_{DD}$. Already using $V_{DD}$ = 0.35V, both CS and axial ideal TFET delays saturate or even degrade (Fig. 9b) so that the axial TFET cannot achieve Si HP delay, even using a larger amount of power. For well-designed TFETs, the best speed performance is typically achieved close to $V_{DD}$ = 0.3V. This is related to the saturation of the TFET current with increasing $V_G$ in the on-state (Fig. 9a). A detailed physical explanation of this effect, which is primarily related to a saturation of the energy windows available for tunneling and secondarily to an increase of the tunneling distance due to source depletion in the on-state, can be found in [2].

## IV. NON-IDEAL CS HTFETs

To verify that the CS TFET steep swing and performance advantage are not lost when fundamental sources of non-idealities are considered, we investigate the impact of electron-phonon scattering (e-ph) and discrete dopant impurities (DI's) on optimized CS nTFETs. Traps are another type of non-ideality that can severely degrade the performance of experimental TFETs [35]. On the contrary to phonons that are intrinsic (at least at non-zero temperature) and related to the lattice vibrations of the material, or DI's that are inherent to the usage of doping, trap concentrations can, in principle, be reduced below a critical level by improving the processing conditions. As such, traps are, therefore, out of the scope of this paper that focus on the fundamental physics and performance of the CS TFET.

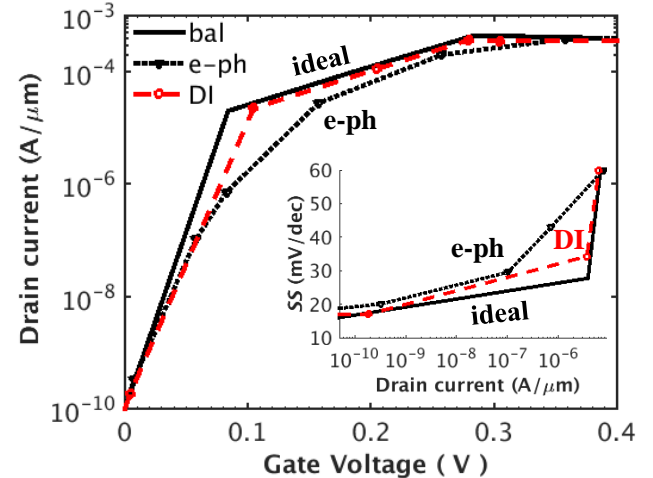

Fig. 10. $I_D(V_G)$ and $SS(I_D)$ (inset) of InAs/GaSb CS NW nTFET ideal (bal), with e-ph, and with DI band tails. $V_D$ = 0.3V. $I_{OFF}$ = 100 pA/µm.

Both e-ph and DI non-idealities may strongly degrade the filtering efficiency and SS of a TFET. Their inclusion in an atomistic framework such as TB NEGF represents the main ingredients for the microscopic treatment of band tails in doped crystalline semiconductor devices [36,37]. Scattering electrons can relax their energy, which potentially increases the probability of final tunneling states and tunneling current but may also increase SS. To tackle the numerical burden of considering e-ph in such large NW devices, we implemented electron-phonon scattering within a mode-space NEGF self-consistent Born approximation and using the efficient form-factor method as described in section II. DI's locally result in a non-uniform spatial potential profile, which in turn yields a spatially varying onset of tunneling. This may degrade the filtering efficiency and SS of the device. DI's are simulated in an atomistic fashion as discrete charges on atomic sites [38] in the core-channel region of the CS device. Different DI configurations, but with a fixed number of DI's corresponding to the target doping concentration, were investigated. For the optimized axial TFETs, we have shown that both non-idealities have a limited impact on $I_D(V_G)$ and performance [2] (Table 1 and Fig. 9b). Fig. 10 shows that the CS device steep swing is more sensitive to e-ph. Due to inelastic collisions with phonons, electrons from the core channel may acquire sufficient energy and make it to the shell, even though $CB_{SH}$ is still at a higher energy than $VB_C$ and the direct (ballistic) BTBT path from the core to the shell is closed (Fig. 4a-b).

At same $I_{OFF}$, due to the degraded slope, a 38% $I_{ON}$ reduction is observed at $V_{DD} = 0.3V$ for the case with scattering. When further increasing $V_G$ (and $V_{DD}$), however, the ideal ballistic current saturates and the non-ideal CS TFET current degradation is only 13% at $V_G = 0.35V$, while it has fully caught-up at $V_G = 0.4V$ with the ballistic current. As a consequence, the optimal delay performance for the non-ideal CS-TFET inverter case is achieved at $V_{DD} = 0.35$ V and is close to that of the ideal case that was achieved at $V_{DD} = 0.3$ V, but with a 1.5 × larger power consumption (Fig. 9b). Overall, the $V_{DD} = 0.35V$ non-ideal CS-TFET inverter delay is comparable to that of the HP Si CMOS using $V_{DD} = 0.55V$ and with a switching energy that is reduced by more than 2×.

V. CONCLUSIONS

Using an efficient atomistic mode-space NEGF technique, we explored the essential physics, design and fundamental performance potential of the III-V line-tunneling core-shell NW HTFET.

The CS TFET "line-tunneling" current increases significantly with the core diameter $d_C$ and outperforms the best III-V axial "point-tunneling" NW heterojunction TFET $I_{ON}$ by up to 6× for $d_C = 6.6$ nm. Reaching such a high level of current at low supply voltage, however, requires and involves specific and sometime unanticipated optimizations and physics (e.g., the need for the shell band alignment that stems from the required large core-channel doping) that were thoroughly investigated here. In spite of the commonly accepted and simpler geometrical view, we also showed and explained the weak gate-length dependency observed for the line-tunneling current in a III-V TFET.

We have further investigated the effect of electron-phonon scattering and discrete dopant impurity band tails on optimized CS NW TFETs. This was enabled by the extension of our atomistic mode-space NEGF algorithm beyond the ballistic approximation. It allowed, for the 1st time, the atomistic simulation of NWs with a diameter as large as 10 nm and including electron-phonon scattering. It was shown that e-ph affects the drive performance of CS TFETs for $V_{DD} < 0.3V$, but that the on-current is quickly recovered for larger $V_{DD}$. As a consequence, the optimal delay performance for the non-ideal CS-TFET inverter case is achieved at $V_{DD} = 0.35$ V and is close to that of the ideal case, that was achieved at $V_{DD} = 0.3$ V, but with a 1.5 × larger power consumption. Overall, the $V_{DD} = 0.35V$ LP non-ideal CS-TFET inverter performance significantly outperforms that of the axial TFETs. The $V_{DD} = 0.35V$ CS inverter delay is comparable to that of the HP Si CMOS using $V_{DD} = 0.55V$ with 100× lower leakage and with a switching energy that is reduced by more than 2×, which shows promise for a low power TFET technology with HP speed.

ACKNOWLEDGMENT

The authors thank Dr. Y. C. Sun for management support.

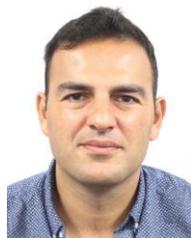

Aryan Afzalian received the electro-mechanical Engineering degree and the Ph.D. degree in electronic engineering from the Université catholique de Louvain (UCL), Louvain-La-Neuve, Belgium, in 2000, and 2006, respectively.

From 2006 to 2009 he was a Postdoctoral Research Fellow at Tyndall National Institute, Cork, Ireland. From 2009 to 2013, he was a senior research associate with UCL, Belgium. Since 2013, he has been with the Research Division of TSMC, located in IMEC, Leuven, Belgium, working on the atomistic modeling of advanced CMOS and beyond CMOS devices.
Dr. Afzalian has over 18 years of experience in semiconductor device simulation and modeling, with specific expertise in quantum transport codes and algorithms (NEGF) and the physics of devices at the nanoscale. He has authored or co-authored more than 100 technical publications in international conferences, journals and books and holds 11 patents. He is the recipient of the 2001 AILV award for his master thesis work on SOI image sensors, and of the 2009 UCC Invention of the year awards for his work on Resonant Tunneling FETs.